\documentclass[a4paper,fleqn,usenatbib]{mnras}

\usepackage{newtxtext,newtxmath}
\usepackage[T1]{fontenc}
\usepackage{ae,aecompl}

\usepackage{graphicx}	% Including figure files
\usepackage{amsmath}	% Advanced maths commands
\usepackage{amssymb}	% Extra maths symbols

\def\arcsec{$^{\prime\prime}$}

\def\na{NewA}
\newcommand{\NtwoH}{N$_{2}$H$^{+}$}
\newcommand{\HCO}{HCO$^{+}$}
\newcommand{\HthrCO}{H$^{13}$CO$^{+}$}
\newcommand{\HthrCN}{H$^{13}$CN}

\newcommand{\Msun}{\mbox{$\mathrm{M}_{\sun}$}}
\newcommand{\Jybm}{Jy~beam$^{-1}$}

\newcommand{\kms}{km~s$^{-1}$}

\title[Filament Formation from Shocked Flows]{Self-gravitating Filament Formation from Shocked Flows:\\
Velocity Gradients across Filaments}

\author[C.-Y. Chen et al.]{
Che-Yu Chen,$^{1}$\thanks{E-mail: cc6pg@virginia.edu}
Lee G. Mundy,$^{2}$
Eve C. Ostriker,$^{3}$
Shaye Storm$^{4}$
and Arnab Dhabal$^{2}$
\\
$^{1}$Department of Astronomy, University of Virginia,
  Charlottesville 22904, VA, USA\\
$^{2}$Department of Astronomy, University of Maryland,
  College Park 20742, MD, USA\\
$^{3}$Department of Astrophysical Sciences, Princeton
  University, Princeton 08544, NJ, USA\\
$^{4}${Harvard-Smithsonian Center for Astrophysics, Cambridge, MA 02138, USA}
}

\date{Accepted XXX. Received YYY; in original form ZZZ}

\pubyear{2019}

\begin{document}

\label{firstpage}
\pagerange{\pageref{firstpage}--\pageref{lastpage}}
\maketitle

\begin{abstract}

In typical environments of star-forming clouds, converging supersonic turbulence generates shock-compressed regions, and can create strongly-magnetized sheet-like layers. 
Numerical MHD simulations show that within these post-shock layers, dense filaments and embedded self-gravitating cores form via gathering material along the magnetic field lines. As a result of the preferred-direction mass collection, a velocity gradient perpendicular to the filament major axis is a common feature seen in simulations. 
We show that this prediction is in good agreement with recent observations from the CARMA Large Area Star Formation Survey (CLASSy),
from which we identified several filaments with prominent velocity gradients perpendicular to their major axes.
Highlighting a filament from the northwest part of Serpens South,
we provide both qualitative and quantitative comparisons between simulation results and observational data. In particular, we show that the dimensionless ratio $C_v \equiv {\Delta v_h}^2/(GM/L)$, where $\Delta v_h$ is half of the observed perpendicular velocity difference across a filament, and $M/L$ is the filament's mass per unit length, can distinguish between filaments formed purely due to turbulent compression and those formed due to gravity-induced accretion. We conclude that the perpendicular velocity gradient observed in the Serpens South northwest filament can be caused by gravity-induced anisotropic accretion of material from a flattened layer.
Using synthetic observations of our simulated filaments, we also propose that a  density-selection effect may explain observed subfilaments (one filament breaking into two components in velocity space) as reported in \cite{2018ApJ...853..169D}.\\
\end{abstract}

\section{Introduction}
% opening paragraph

Filaments are prevalent in observed star-forming clouds
\citep{1979ApJS...41...87S,1987ApJ...312L..45B,2008ApJ...680..428G}, and are generally considered to be connected with dense, star-forming cores \citep{2010A&A...518L.102A,2010A&A...518L.106K,2013ApJ...777L..33P}.
% review of observations
Since the extensive studies completed with {\it Herschel} \citep{2010A&A...518L.104M,2010A&A...518L..92W}, a number of properties of observed filaments have been intensively studied including the density and temperature profiles (\citealt{2011A&A...529L...6A,2019A&A...621A..42A,2012A&A...541A..12J,2013A&A...550A..38P}, or see \citealt{2014prpl.conf...27A} for a review), the velocity sub-structures termed ``fibers" \citep{2013A&A...554A..55H}, and the relative orientation between filaments and magnetic fields as revealed by {\it Planck} \citep{PlanckXXXV}.
In addition, molecular line observations with dense gas tracers have also been used to probe the detailed kinematics within individual filamentary systems \citep{2013ApJ...766..115K,2014ApJ...790L..19F, 2018ApJ...853..169D}.

% review of filament formation theory
In theoretical studies, filamentary structures appear in large-scale simulations investigating turbulence-induced cloud evolution either with or without magnetic fields and/or self-gravity
\citep{1999ApJ...513..259O,2001ApJ...546..980O,2004ASPC..322..299K,2005A&A...435..611J,2007ApJ...661..972P,2008ApJ...687..354N,2016MNRAS.457..375F},
and also in more idealized studies of gravitational/magnetic instabilities in sheet-like clouds
\citep{2006ApJ...652..442C,2009NewA...14..221B,2014ApJ...789...37V},
While filamentary structures can arise from a variety of dynamical processes, the densest ones, which ultimately host the formation of stars, are likely to form in gas that has been strongly compressed by a converging, supersonic flows.  Numerical studies of colliding flows which also include turbulence have indeed demonstrated the formation of filaments in post-shock layers and creation of embedded self-gravitating cores 
\citep{2007ApJ...657..870V,2011ApJ...729..120G,2014ApJ...785...69C,2015ApJ...806...31G}.
Observations of filament separation in clouds has also suggested that
filaments form within shock-compressed sheets
\citep{2002ApJ...578..914H}.

As filaments are seen in a diversity of environments, and at a range of scales, they may have several different formation mechanisms. 
Filaments of cold gas in the warm ISM may  occur as a result  of thermal instability in combination with turbulent compression and shear \citep[e.g.,][]{2005A&A...433....1A,2005ApJ...629..849P,2006ApJ...648.1052H,2006ApJ...643..245V,2009ApJ...704..161I}. Within entirely-molecular gas, filaments are commonly considered to be the products of  direct compression of interstellar  turbulence, or as part of the cloud-scale gravitational collapse \citep[see e.g.,][]{2013A&A...560A..68H,2014ApJ...791..124G,2016ApJ...831...46A}.
Though filaments are often seen to have velocity substructure in both
observations and simulations
\citep[e.g.][]{2013A&A...554A..55H,2015ApJ...807...67M}, the classical
filament model of a static, self-gravitating, infinitely long cylinder
\citep{1964ApJ...140.1056O} is still widely adopted to interpret
observation results
\citep[e.g.][]{1999ApJ...510L..49J,2005A&A...440..151H,2011A&A...529L...6A}.
Static filament models with non-zero external pressure are more
realistic than filaments of infinite radius for comparison to real
clouds \citep{2012A&A...542A..77F}. It is also possible to
consider temporal evolution of self-gravitating cylinders in various
limits
\citep[e.g.][]{1992ApJ...388..392I,2013ApJ...776...62H}. However, the lack of symmetry in molecular clouds and their dynamic nature implies that filaments in general must form from gas structures that are not cylindrically symmetric \citep{2013ApJ...769..115H}, and signatures of the formation may be evident in the velocity fields around filaments. In particular, when filaments form via self-gravitating contraction in a shock-compressed dense layer, the velocity field surrounding the center of the forming filament will have converging-flow motions primarily in a plane containing the filament (\citealt{2014ApJ...785...69C,2015ApJ...810..126C}; hereafter \hyperlink{CO14}{CO14}, \hyperlink{CO15}{CO15}).

% new findings in this paper
In this paper, we describe the kinematic properties of forming filaments as obtained in numerical MHD simulations, and compare to kinematic features revealed in molecular line observations conducted by the Combined Array for Research in Millimeter-wave Astronomy (CARMA) towards nearby star-forming clouds, as part of the {\it CARMA Large Area Star Formation Survey} (CLASSy) project \citep{2014ApJ...794..165S,2016ApJ...830..127S,2014ApJ...797...76L}.
Several filaments in the Perseus and Serpens Molecular Clouds
show clear gradient across their major axes in line-of-sight velocity \citep{2014ApJ...790L..19F, 2018ApJ...853..169D},
a signature feature of preferred-direction accretion described in
\hyperlink{CO14}{CO14}. 
%In addition, the relatively uniform thermal line widths 
In addition, the relatively narrow turbulent line widths at high angular resolution
in Perseus and Serpens indicate that these are
flattened structures along the line of sight with depth $\sim 0.3$~pc
\citep{2014ApJ...797...76L,2014ApJ...794..165S}, in agreement with the
expected thickness of the post-shock layer created by converging flows (e.g.~\hyperlink{CO14}{CO14}, \hyperlink{CO15}{CO15}).  We
therefore suggest that these observed filaments in Perseus and Serpens
are forming via preferred-direction accretion within locally flat regions, which themselves may have
been generated by large-scale shocks within each cloud.

The outline of this paper is as follows. 
We review and illustrate from simulation data our proposed model of filament formation within shocked layers in Section~\ref{sec::theory}.
In Section~\ref{sec::sim} we describe the numerical simulations we employ from \hyperlink{CO14}{CO14} and \hyperlink{CO15}{CO15}, 
as well as how we characterize the filaments (Section~\ref{sec::filaId}). Section~\ref{sec::evo} presents discussions of the key properties of filaments identified in these simulations, including an evolutionary study and the quantitative measurements of gravity-induced velocity gradient. We provide a detailed comparison to observational data from CLASSy in Section~\ref{sec::obs}, where we also discuss a possible origin of multiple velocity components within individual filaments (Section~\ref{sec::density}).
We summarize our conclusions in Section~\ref{discussion}.

\section{Filament Formation within Flat Layers: Kinematic Signatures}
\label{sec::theory}

\begin{figure}
\begin{center}
\includegraphics[width=\columnwidth]{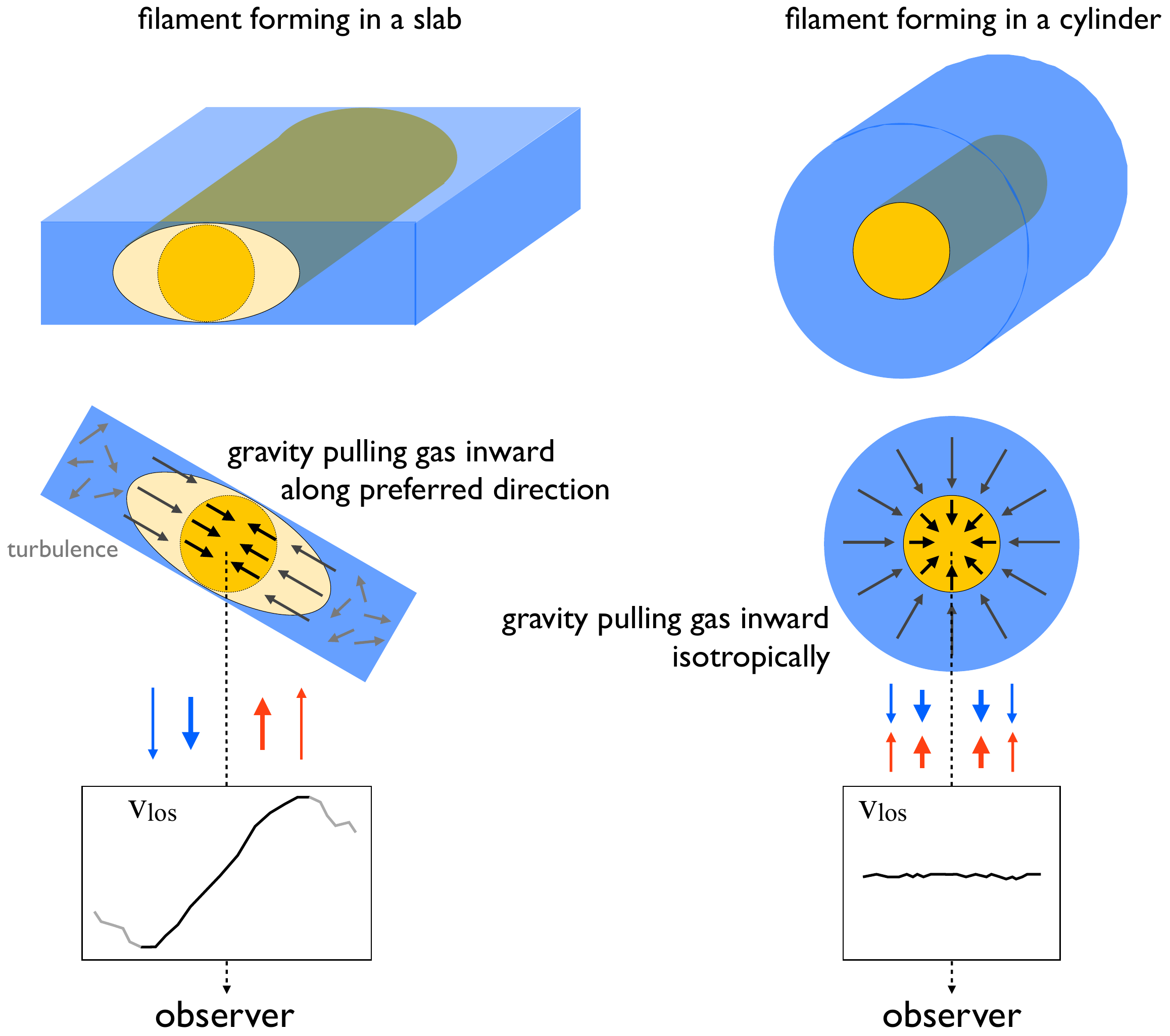}
\vspace{-.15in}
\caption{The ``preferred-direction" filament formation versus
  isotropic gas flow. The observed velocity gradient across the filament indicates there is a preferred direction for gas flows
  near the filament (\textit{left}). Filament formation within 
  a compressed layer naturally exhibits convergent flow within the layer which appears as a local velocity gradient on the plane of the sky. In contrast,
  if the filament formation is cylindrically-symmetric, material is drawn in equally from around the cylinder and there is no gradient along the line-of-sight velocity, since the
  red-shifted and blue-shifted components are symmetric
  (\textit{right}). }
\label{formation}
\end{center}
\end{figure}

%In simulations of converging flows, filament-like structures are commonly seen in the shock-compressed layer (see e.g.~Figure~6 in \hyperlink{CO14}{CO14}).
%Within these layers, small-scale perturbation initiates local overdensities. 
In simulations of large-scale supersonic converging flows, small-scale perturbation initiates local overdensities within the shock-compressed layer.
%With the density of the layer growing slowly, the turbulence in the layer then takes the "almost" self- gravitating lay and imposes the filamentary structure onto it by creating a seed which grows in density faster than the layer.
In the case that the large-scale 
converging flow produces a shocked layer that becomes marginally self-gravitating,
%collision is strong enough to drive the shocked layer to the edge of self-gravity,
even small velocity perturbations would induce formation of overdense structures that start gravitationally
%Self-gravity enhances these perturbations by attracting surrounding gas, 
pulling in material to form added filaments and dense cores.
Such two-stage model for core formation in strongly magnetized layers is described in \hyperlink{CO14}{CO14} and \hyperlink{CO15}{CO15}.  

Since the perpendicular component of the velocity is mostly lost in transitioning through the shock,
%Since the thin post-shock layer confines the vertical movement of the shocked gas, 
the gas velocity within the compressed layer is mostly parallel to the plane of the shock front.
Therefore, if viewed from an angle not perfectly face-on to the
post-shock layer, the line-of-sight velocity around a filament will
have a gradient across its major axis. This ``preferred-direction" filament formation process is illustrated in Figure~\ref{formation}. The gradient is initially due to the relatively weak, locally-converging velocity perturbation within the post-shock layer that is necessary to generate the seed of the filament. This proto-filament grows faster than the layer, and once it becomes overdense enough to be strongly self-gravitating, it draws in the denser infalling material from the post-shock layer which maintains and enlarge the velocity gradient.

Note that in the filament formation scenario described above, magnetic field is not required for filaments to show velocity gradients perpendicular to their major axes (see e.g. Fig.~2 of \citealt{2011ApJ...729..120G}). 
However, in the presence of dynamically-important magnetic field, the preferred-direction mass accretion guided by magnetic field lines could enhance the detection of velocity gradients across filaments by regulating gas flows surrounding the filaments.
Also note that, though the flows that produce filaments are primarily along magnetic fields, it is not necessary for filaments to be strictly perpendicular to the local magnetic field, because the loci of maximum density along each filament need not be exactly perpendicular to the magnetic field direction (see e.g. Figure~\ref{cloudEvo} below).
%For the example shown in Figure~\ref{gradientSample}, the magnetic field in the post-shock layer runs diagonally from lower-left to upper right.

We emphasize that our proposed model of filament formation differs from that described in \cite{2016ApJ...831...46A}, which considered filaments as ribbon-like structures that are the direct products from magnetic field-regulated turbulent  compression. 
%One critical point on understanding the forming process of filament is that, since filaments are single-dimensional, string-like structures while the colliding turbulent cells are 3D, the formation of filaments must be a 2-step process;  i.e., the turbulent cells collide and compress gas in between to form a  dense  layer  (a  2D  structure)  first,  then  filaments  form within this locally-flat region. 
In our picture, 
%filaments are one-dimensional, string-like structures, while the colliding turbulent cells are relatively isotropic in three-dimensional space. 
the formation of filaments is
%such filaments therefore must be 
a two-step process:
%because there is no natural way to form a one-dimensional structure directly from three-dimensional configurations: 
a converging portion of a supersonic turbulent flow first compresses gas to form a dense layer (a two-dimensional structure), and then 
the one-dimensional, string-like
filaments form within this locally-flat region via anisotropic gas accretion as shown in Figure~\ref{formation}.
This filament formation scenario also differs from that investigated in \cite{2013A&A...560A..68H} and \cite{2014ApJ...791..124G}, which formed filaments via the gravitational collapse of a molecular cloud as a whole. 
One would not expect to see velocity gradient across the major axis of the filament under such global contraction, which is more  similar to the case shown in the right panel of Figure~\ref{formation}.

\begin{figure*}
\begin{center}
\includegraphics[width=0.8\textwidth]{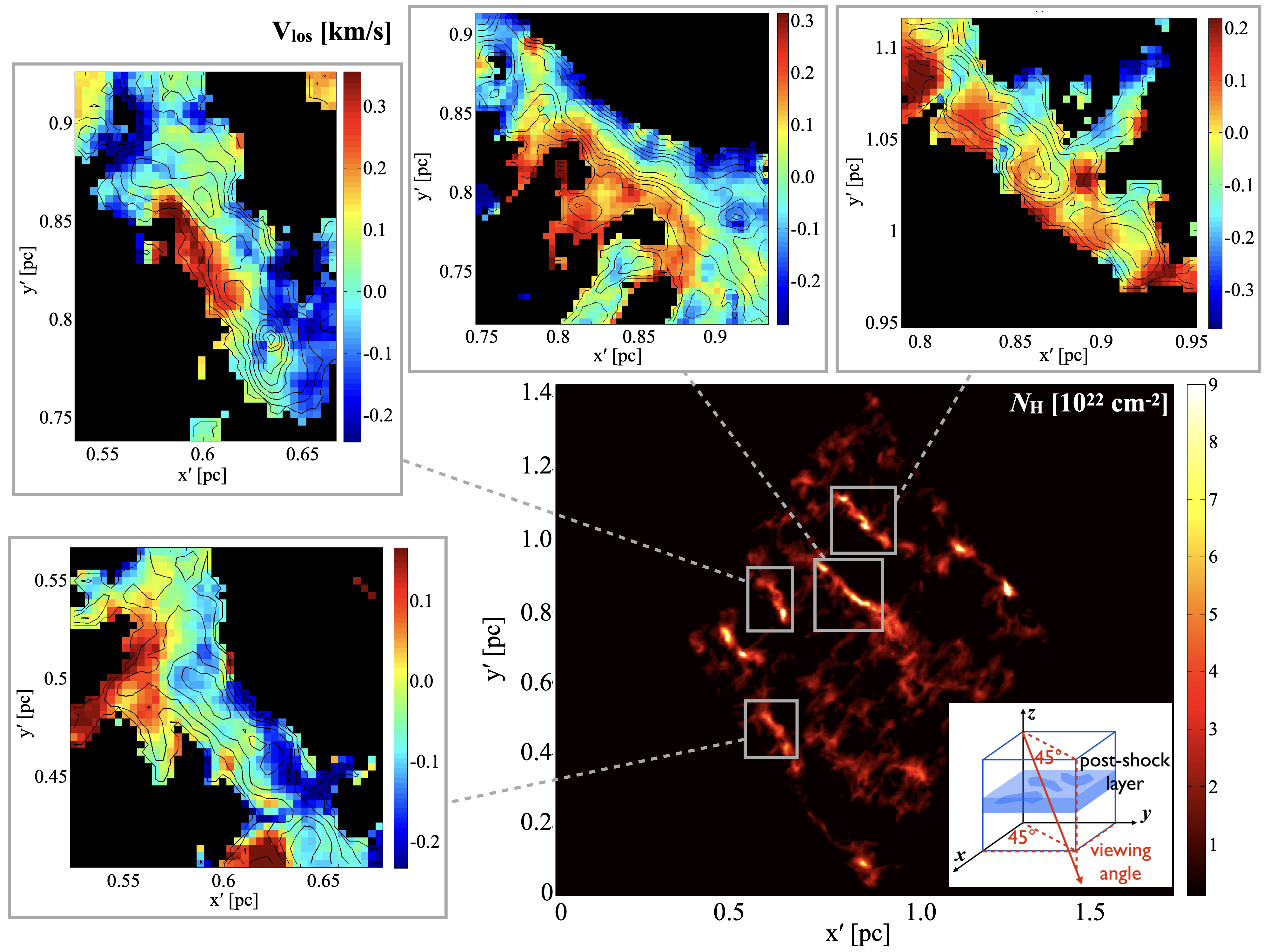}
%\vspace{-.15in}
\caption{The kinematic features of filaments that form in self-gravitating converging turbulent flow simulations. In the example simulation from \protect\hyperlink{CO14}{CO14}, large-scale supersonic turbulence converges locally along the $z$-direction, forming a post-shock layer in the $x$-$y$ plane (\textit{insert, bottom-right}). When viewed from any direction (except perfectly edge-on), the integrated column density map (\textit{bottom-right}) shows filamentary structures formed within the dense post-shock layer. Loci of four line-of-sight velocity maps are marked with gray boxes. A velocity gradient perpendicular to the major axis of each filament is a common feature for filaments formed within a dense layer when viewed at any angle except perfectly face-on. }
\label{gradientSample}
\end{center}
\end{figure*}

Figure~\ref{gradientSample} illustrates some examples of this kind of velocity gradients within filaments described above, from synthetic observations of model A5ID of \hyperlink{CO14}{CO14}. The column density and the density-weighted line-of-sight velocity are projected along the direction that is $45^\circ$ from the post-shock plane and $45^\circ$ from the rough direction of the mean magnetic field in the layer (approximately along $\hat{\mathbf{x}}$), which is roughly perpendicular to the largest filaments (see the insert at bottom right on Figure~\ref{gradientSample}). For closer comparison to observations using a dense-gas tracer (e.g.~the CLASSy data discussed below in Section~\ref{sec::obs}), we only include voxels from the computational output with number densities between $10^4-10^7$~cm$^{-3}$.
Figure~\ref{gradientSample} demonstrates that prominent velocity gradients across the filaments' major axes are commonly seen in simulated star-forming regions generated by shocks, which could be an observable feature of filaments formed within a flattened layer.

However, gravity-induced anisotropic accretion is not the only way to produce a velocity gradient perpendicular to the filament's major axis. Filaments formed from direct compression of local shocks could also show such velocity gradients. 
These filaments are likely confined by external ram pressure provided by the shock flows, which would result in a more significant velocity difference across the filament. 
Here, we propose that these two scenarios of filament formation could be easily distinguished quantitatively if the velocity difference across the filament and the mass per unit length of the filament are known.
%by comparing the velocity difference across the filament to its gravitational potential, to see if the self-gravity of the filament is strong enough to generate such velocity difference. 

For simplicity, we consider an idealized cylindrical filament with mass $M$ and
length $L$. For radial contraction induced by self-gravity, we have
\begin{equation}
\ddot{r} = -2\frac{GM(r)/L}{r}.
\end{equation}
This can be integrated to give \citep[see e.g.][]{2009ApJ...704.1735H}
\begin{equation}
\frac{1}{2}\dot{r}^2 = 2G\ln\left(\frac{r_0}{r}\right)M(r)/L,
\label{rdot}
\end{equation}
which represents the radial velocity of gas being pulled gravitationally from distance $r_0$ to $r$ by the filament.
The ratio $r_0/r$ can be approximated as $r_0/r \sim \Sigma/\Sigma_0$ by considering the mass per unit length of the filament, $M/L = \Sigma \cdot r \sim \Sigma_0\cdot r_0$, where $\Sigma$ is the column density of the filament, and $\Sigma_0$ represents the average column density of the surrounding environment wherein the filament has formed.
It has been shown in \hyperlink{CO15}{CO15} that $\Sigma/\Sigma_0\sim 2-20$
for sufficiently advanced times such that the filament becomes prominent (see their Figure~6), which means the logarithm term in Equation~(\ref{rdot}), $\ln (r_0/r)$, can be considered as order of unity.
Equation~(\ref{rdot}) therefore would yield
\begin{equation}
\dot{r}^2 \sim G M(r)/L
\label{rdot2}
\end{equation}
for gravitationally contracting filaments.

As discussed above, the velocity gradient across a filament may be
either induced by the self-gravity of the filament, or may simply reflect supersonic turbulence of the cloud. To quantitatively distinguish
between these two scenarios, we define a dimensionless coefficient
$C_v$ to compare the ratio between the kinetic energy of the flow transverse to the filament and the gravitational potential energy of the filament gas:
\begin{equation}
C_v \equiv \frac{{\Delta v_h}^2}{GM(r)/L}.
\label{coeff}
\end{equation}
Here, $\Delta v_h$ is half of the velocity difference across the
filament out to a transverse distance $r$ (on both sides), and $M(r)/L$ is the mass
per unit length of the filament measured at the same distance from the
spine as the $\Delta v_h$ measurement.  The value of $C_v$ is
suggestive of the origin of the velocity gradient.  If $C_v \gg 1$,
the local turbulence is much stronger than the filament's
self-gravity, and the filament is likely forming as a result of
shock compression. If $C_v \lesssim 1$, the gravitational potential energy is comparable to the gas kinetic energy, which following
Equations~(\ref{rdot}) or (\ref{rdot2}) suggests that the velocity
structure is likely induced by the filament's self-gravity.
We would like to point out that filaments could in principle have $C_v \ll 1$ because of projection effects, or if a filament is at an early formation stage in our proposed scenario. Slowly re-expanding filaments could also have $C_v \ll 1$; however, we note that these filaments are shorter-lived and are less likely being seen in observations.

Below we provide examples and tests of our filament formation model using both numerical simulations and existing observation data.

\section{Filaments in Simulations}
\label{sec::sim}

\begin{figure*}
\begin{center}
\includegraphics[width=\textwidth]{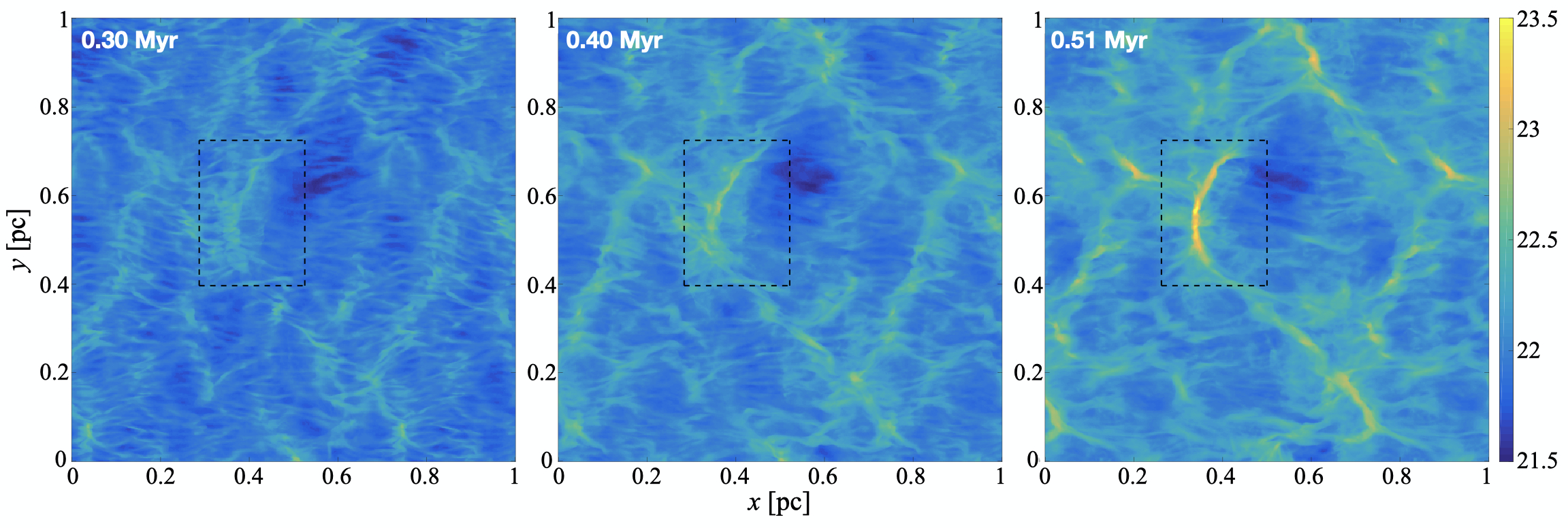}
\vspace{-.2in}
\caption{Snapshots of gas column density showing evolution to create filaments in model M10B10 from \protect\hyperlink{CO15}{CO15}. The post-shock magnetic field lies parallel to the $\hat{x}$ direction.  Gravity causes gas to flow along the magnetic field towards local density maxima, and over time filaments build up from accreted material.  The inset box in dashed lines marks the region analyzed in Figure~\ref{filaEvo} below.}
\label{cloudEvo}
\end{center}
\end{figure*}

\subsection{Simulations}

The simulations we use are reported in \hyperlink{CO14}{CO14}, \hyperlink{CO15}{CO15} and summarized here. Those simulations, considering a magnetized shocked layer produced by plane-parallel converging flows, were conducted using the \textit{Athena} MHD code \citep{2008ApJS..178..137S}, with box size $(1~\mathrm{pc})^3$ and resolutions $256^3 - 512^3$ (such that $\Delta x \approx 0.002-0.004$~pc).  
In these simulations, box-scale supersonic inflows (${\cal M}=10$; modeling the largest-scale turbulence in a cloud) collides head to head along the $z$-axis of the simulation box, creating a flat post-shock region in the $x$-$y$ plane.  The flow includes a local perturbed velocity field that follows the scaling law for turbulence in GMCs $\sigma_v (\ell)\propto \ell^{1/2}$ (with a Fourier power spectrum $v^2(k)\propto k^{-4}$; see \cite{2011ApJ...729..120G}), with a largest scale of $1/2$ of the box size. For simplicity, an isothermal equation of state with sound speed $0.2$~km/s is adopted.  
%Though \hyperlink{CO14}{CO14} included both non-ideal and ideal MHD investigations, we only adopt those with ideal MHD for our use in this contribution. 
We adopt model M10B10 from \hyperlink{CO15}{CO15} for our use in this work except in Figure~\ref{gradientSample}, which uses model A5ID from \hyperlink{CO14}{CO14} to illustrate the kinematic feature because there are more separated filaments formed in this particular simulation.

The initial magnetic field in the simulation is set to be oblique to the shock on the $x-z$ plane with total magnitude $10~\mu$G. 
Within the post-shock layer, the magnetic field component parallel to the layer ($B_x$) is strongly enhanced by compression, while the perpendicular component ($B_z$) is not.  
The post-shock magnetic field is therefore relatively well-ordered, approximately along the $x$ direction.
Though the exact field strength depends on the angle between the initial magnetic field and the inflow, the post-shock magnetic pressure is approximately equal to the ram pressure in the inflow in the strong shock limit: ${B_\mathrm{ps}}^2/(8\pi) \sim \rho_0 {v_0}^2$ (see \hyperlink{CO14}{CO14} and \hyperlink{CO15}{CO15}). This applies when the post-shock region is dynamically dominated by the shock-amplified magnetic field.

As discussed in Section~\ref{sec::theory}, to observe the ``preferred-direction" filament formation as illustrated in Figure~\ref{formation}, the post-shock layer must have a non-zero relative angle with respect to the plane of sky (see e.g.~Figure~\ref{gradientSample}). Therefore in the analysis discussed below, the simulation box has been rotated by $\theta = 45^\circ$ around the $y$ axis. We note that though this viewing angle affects the measured velocity difference across the filament (and therefore the value of $C_v$), it is not critical in our analysis as long as it is not the extreme cases ($\theta \sim 0^\circ$ or $\theta \sim 90^\circ$). Some discussions of the viewing angle effect are included in Section~\ref{sec::density}.

Figure~\ref{cloudEvo} shows an example of the evolution of gas structures formed in the post-shock region (in column density) at $45^\circ$ viewing angle; since the simulation box is periodic in the $x$ direction, there are repeated patterns near the edge of the box.
Filamentary structures become prominent at a time $\sim 0.3-0.5$~Myr after the initial collision of the convergent flows, with major (more massive) filaments aligned roughly perpendicular to the local magnetic field.\footnote{There are also thin, hair-like sub-filaments (or striations) that seem to follow the direction of magnetic field; these are not our main focus in this study. For further discussions regarding these field-aligned striations, see \cite{2017ApJ...847...140C}.}
We can already see the footprints of anisotropic gas accretion onto the filaments from these sequential pc-scale maps; in the next section, we focus on the $\sim 0.3\times0.2$~pc$^2$ zoom-in region around the main filament of this simulation (marked by the dashed-line box in Figure~\ref{cloudEvo}) for quantitative analysis.

%need to identify individual filaments and define their boundaries. 

\subsection{Characterizing Filaments}
\label{sec::filaId}

\begin{figure*}
\begin{center}
\includegraphics[width=\textwidth]{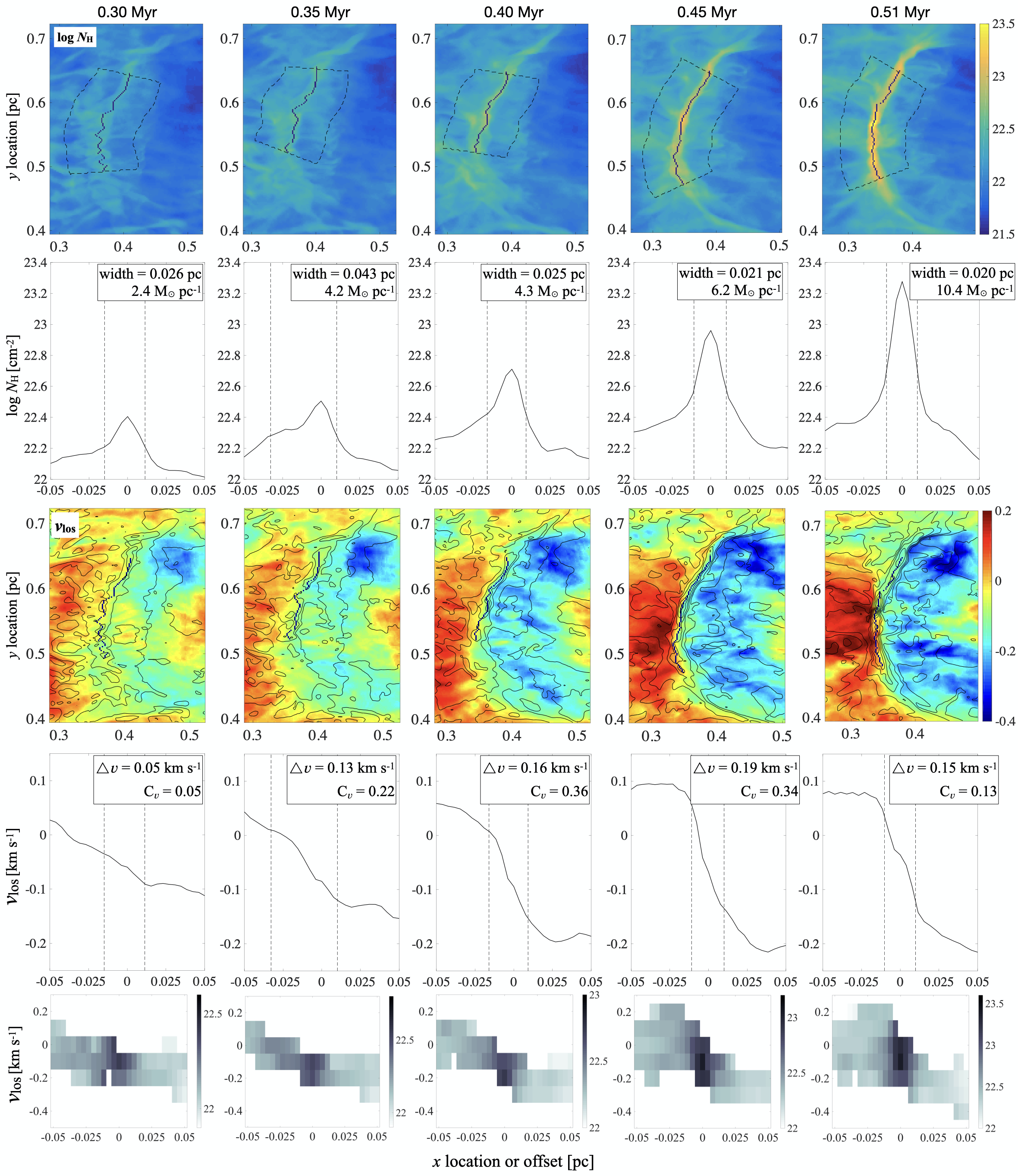}
\vspace{-.1in}
\caption{The evolution of a sample filament (marked in dashed box in Figure~\ref{cloudEvo}), showing the $N_\mathrm{H}$ column density map ({\it top row}, in $\log$ scale) and profile ({\it second row}), the line-of-sight velocity ({\it third row}, in km~s$^{-1}$) and profile ({\it fourth row}), and the converted position-velocity diagram ({\it bottom row}).
The dashed regions in the column density maps mark the $0.05$~pc radius from the spine ({\it black dots}) that are considered in the profile plots and PV diagrams. The boundaries of each filament (as defined in Section~\ref{sec::filaId}) are marked by vertical dashed lines in either the column density or velocity profile plots. The filament width, mass per unit length, velocity difference across the filament, and the $C_v$ coefficient are also noted in those panels. }
\label{filaEvo}
\end{center}
\end{figure*}

%{\bf <<how I identify the ``spine" of the filament and how to define the filament boundary from column density profile; tbc>>}
As filaments are effectively projected structures in 2D in observations, we use the column density map to define filaments. For the main filament in this simulation and within the zoom-in region from Figure~\ref{cloudEvo},
we picked the two ends of the filament that we want to analyze, and then marked the ``spine" of the filament by finding the maximum column density at each row of $y$. We then calculate the distance to the spine for each pixel, and consider only pixels at distance $d_\mathrm{fila}<0.05$~pc to the spine in our following analysis. 
These are shown in the top row of Figure~\ref{filaEvo}, where we over-plotted the spine of the filament and the $\pm 0.05$~pc area on the column density map of the zoom-in region from Figure~\ref{cloudEvo}. 
Note that this selection of $d_\mathrm{fila}$ takes the recent {\it Herschel} results (that typical filament width $\sim 0.1$~pc) into consideration, and also (based on the results) is large enough to cover the velocity gradients across the filaments formed in this simulation.\footnote{In fact, note that the simulation considered here (model M10B10 from \hyperlink{CO15}{CO15}) is an ideal MHD simulation, and is therefore scale-free, which means it can be easily rescaled to represent a different physical scale \citep[see e.g.][]{2018MNRAS.474.5122K}. This is why we do not discuss, or compare with those measured in observations, the width of simulated filaments here.}

We can therefore plot the column density and line-of-sight velocity profiles of the filament with respect to offset from the spine, which are shown in the second row (column density profile) and fourth row (velocity profile) of Figure~\ref{filaEvo}.
Using the column density profile, 
we chose to define the filament width in a similar way as the full width at half maximum (FWHM); i.e.~we label the edges of the filament based on where the column density (in $\log$ space) drops to $50\%$ of (maximum$-$background) above the background. Here, the ``background" column density is defined as the minimum value of the column density profile, which is calculated within $\pm d_\mathrm{fila}$ from the spine.
These filament boundaries are marked as vertical dashed lines in both the column density and line-of-sight velocity profiles in Figure~\ref{filaEvo}, with the filament width labeled on the top-right corner of the column density plot (second row) at each time frame.
We note that, though our method of characterizing filaments and the choices of parameter values may seem artificial, it does not affect our results on filament kinematics significantly, since we focus on relative comparisons among evolutionary sequences instead of exact measurements of physical properties at a specific time. 
In addition, as most of those filament parameters (mass per unit length, width, etc.) are highly uncertain in observations, we are only making order-of-magnitude comparisons, and therefore the detailed modeling of specific filaments is not the main focus here.

\subsection{Filament Kinematics and Evolution}
\label{sec::evo}

%Figure~\ref{filaEvo} shows the evolution of the filament marked in the dashed box in Figure~\ref{cloudEvo}. {\bf <<tbc>>}
%To derive the mass per unit length and the $C_v$ coefficient, filament boundaries are needed. 

The mass per unit length of the filament can be derived by integrating filament column density over the filament width, which is labeled on the top-right corner of each column density profile in Figure~\ref{filaEvo} (second row). Not surprisingly, though the filament width does not vary much over the $\sim 0.2$~Myr time coverage of Figure~\ref{filaEvo}, the mass per unit length increases by almost a factor of 5. 
%At later time ($\sim 0.4-0.51$~Myr), the nearly-constant filament width could be explained by the fact that the inflows are supersonic, and thus provide ram pressure to spatially confine the filament. However, between $\sim 0.3-0.4$~Myr,
The nearly-constant filament width could be a hint that the filament is already self-gravitating at $\sim 0.3-0.4$~Myr, because otherwise the filament width should grow when materials flow in and accumulate, thus increase with its mass.
This agree with our filament formation model described in Section~\ref{sec::theory}, that the post-shock layer was already at the verge of gravitational instability when the filament started forming.

Using the column density-defined filament boundaries, we calculated the velocity difference $\Delta v$ across the filament and the corresponding $C_v$ coefficient, both labeled on the top-right corner of the velocity profiles in Figure~\ref{filaEvo} (fourth row). 
Combining with the maps of density-weighted average of line-of-sight velocity in Figure~\ref{filaEvo} (third row), we see clear evidence of gravity-induced accretion onto this filament.
At $\sim 0.30$~Myr after the shock compression (the first time frame of Figure~\ref{filaEvo}), the right-hand-side of the filament does not show clear accretion flows onto the filament.\footnote{Apparently, the local turbulence at the left-hand-side of the filament happens to be originally strong enough and pushing gas towards the filament, which could be the reason of the formation of the ``seed" of this filament in the first place. } This is also reflects in the relatively-gradual slope of the velocity profile and small $\Delta v$ ($0.05$~km~s$^{-1}$).
As the filament becomes more and more massive ($0.35-0.40$~Myr), inward movements towards the filament spine start to emerge on the right-hand-side of the filament. This acceleration of gas can be seen on the velocity profile plot that $v_\mathrm{los}$ becomes more negative at positive offsets, and the increasing velocity difference across the filament ($\Delta v = 0.16$~km~s$^{-1}$ at $0.40$~Myr). 

Interestingly, 
at the time frame $0.45$~Myr, the gas motion and the filament self-gravity seem to reach a balance, as the gas velocity on the left-hand-side of the filament (but outside the filament boundary) remains roughly constant (see the velocity profile plot on the fourth row of Figure~\ref{filaEvo}). 
Though we cannot say the same for the right-hand-side (the side with positive offsets) of the filament, when comparing the velocity profiles at $t=0.45$~Myr and $0.51$~Myr (the last two columns in Figure~\ref{filaEvo}) we see that the right-hand-side gas velocity did not change much over time, indicating that a rough balance between gas motion and the filament's self-gravity has also been reached. 
We note that the last time frame, $t=0.51$~Myr, in Figure~\ref{filaEvo} marks the onset of gravitational collapse of the dense core within the filament;\footnote{As described in \hyperlink{CO14}{CO14} and \hyperlink{CO15}{CO15}, we define the onset of gravitational collapse of a dense core as when its maximum density reaches $\gtrsim 10^7$~cm$^{-3}$.} this is reflected on the most central part of the velocity profile, where a transition from a smooth slope (as in $t=0.45$~Myr) to a relatively flat curve (i.e.~constant velocity) happens. This flat region could be a result of radial collapse of the filament. 

The evolution of velocity structure during the formation of the filament can be summarized by the position-velocity (PV) diagrams shown in Figure~\ref{filaEvo} (last row). We clearly see that the distribution of gas in the velocity space is initially flat (small $\Delta v$; $t=0.30$~Myr), and the slope becomes steeper and steeper when the filament accretes more material, forming a bright hub at the center of the PV space ($t=0.35-0.45$~Myr). 
After $\sim 0.45$~Myr, the balance between the filament's self-gravity and surrounding gas motion is reached, and therefore the maximum gas speed on each side of the filament varies little over time. More interestingly, at $t=0.51$~Myr we see the PV distribution of intermediate-density gas extends towards the reference velocity (i.e.~the velocity of the filament spine at offset $=0$) and is less concentrated (spreads out to larger offsets), indicating the end of gravity-induced accretion of gas. 

%The maximum velocity difference across the filament is reached when %the filament's self-gravity reaches a balance with the surrounding gas, which can be shown by comparing the steepest slopes of gas distributions on the PV diagrams between $t=0.45$~Myr and $t=0.51$~Myr. The radial collapse of the filament can also be hinted by comparing the PV locations of gas with intermediate column densities between these time frames, because they moved from more positive/negative velocities to within similar velocities of the filament spine. 

Finally, as discussed in Section~\ref{sec::theory}, we calculated the dimensionless coefficient $C_v$ as a quantitative estimate of the relative importance between gas kinetic energy and the filament's self-gravity. The value of $C_v$ at each evolutionary step is provided on the top-right corner of the velocity profile in Figure~\ref{filaEvo} (fourth row). One can immediately note that $C_v \sim 0.1-0.3$ over the forming process of this filament, which demonstrates that the filament's self-gravity is playing a critical role in shaping the velocity profile around this filament. 

\section{Filaments in Observations}
\label{sec::obs}

Prominent velocity gradients perpendicular to the filament major axes have been observed in observations, in both a filamentary infrared dark cloud \citep{2015A&A...584A..67B} and nearby star-forming regions \citep{2013A&A...550A..38P,2018arXiv181106240S}. Those velocity features are, in a number of cases, very similar to what is seen in our preferred-direction mass accretion model when the filaments form within locally-flat regions (see Section~\ref{sec::theory}). 
%Recently, the CARMA Large Area Star Formation Survey (CLASSy) observed five star-forming regions in the Perseus and Serpens Molecular Clouds to investigate the structure and kinematics of dense gas in nearby, clustered environments. 
%Several filaments identified in CLASSy data show prominent velocity gradients perpendicular to their major axes, and 
Here, we highlight a filament from the northwest part of the Serpens South Molecular Cloud to conduct quantitative analysis and compare with our numerical models.

\subsection{Observations: CLASSy and CLASSy-II}

The data we use are from the CARMA Large Area Star Formation
Survey (CLASSy) and follow-up observations (referred as CLASSy-II). 
The CLASSy project spectrally imaged \NtwoH{}, \HCO{}, and HCN ($J=1\rightarrow0$) over 800 square arcminutes of the Serpens and Perseus Molecular Clouds, focusing on the NGC\,1333, Barnard~1, and L1451 regions within Perseus, and the Main and South regions of Serpens \citep{2014ApJ...794..165S,2016ApJ...830..127S,2014ApJ...797...76L}. 
%The full details of CLASSy observations, data calibration, and spectral imaging are published in \cite{2014ApJ...794..165S} and \cite{2014ApJ...797...76L} 
%with an analysis of the Barnard~1 region of Perseus and Serpens Main, respectively. 
%The overview of CLASSy project and data calibration are published in \cite{2014ApJ...794..165S}, and the CLASSy-II data is firstly reported in \cite{2018ApJ...853..169D}.
The observational details are given in those papers; relevant points for this discussion are that the velocity resolution was $\sim0.16$~\kms{} and the spatial resolution was $\sim 7.6$\arcsec.
%The observing was done using CARMA23 mode, which cross-correlates all 23 CARMA antennas, split evenly between the two most compact array configurations, resulting in a synthesized beam $\sim$7\arcsec{}. 
%We simultaneously observed in single-dish mode during tracks with stable atmospheric opacity to recover autocorrelation data, and used the autocorrelations from 10.4-m CARMA dishes to recover the spatial scales missing from the interferometric-only data. 
%Each spectral line was observed in a 8~MHz band, providing velocity resolution $\sim$0.16~\kms. The rms of the science-ready spectral images are $\sim$0.13~\Jybm{}, or $\sim$0.3~K brightness temperature, with some variation between CLASSy regions and individual lines.
%We have discovered velocity gradients across the width of several filaments in our CLASSy sample. {\bf review Arnab+18 paper}

From the CLASSy sample, several filaments with velocity gradients across the filament width are detected \citep[see e.g.][]{2014ApJ...790L..19F}; the CLASSy-II project \citep{2018ApJ...853..169D} followed up five filaments detected in CLASSy samples to further investigate the kinematics of filamentary structures. As a test of whether the observed velocity features arise from  a   chemical effect, CLASSy-II adopted optically thin dense gas tracers \HthrCO{} and \HthrCN{} that were not included in CLASSy. More details about the CLASSy-II observations can be found in \cite{2018ApJ...853..169D}.

\subsection{The Serpens South NW (SSNW) Filament}

\begin{figure*}
\begin{center}
\includegraphics[width=\textwidth]{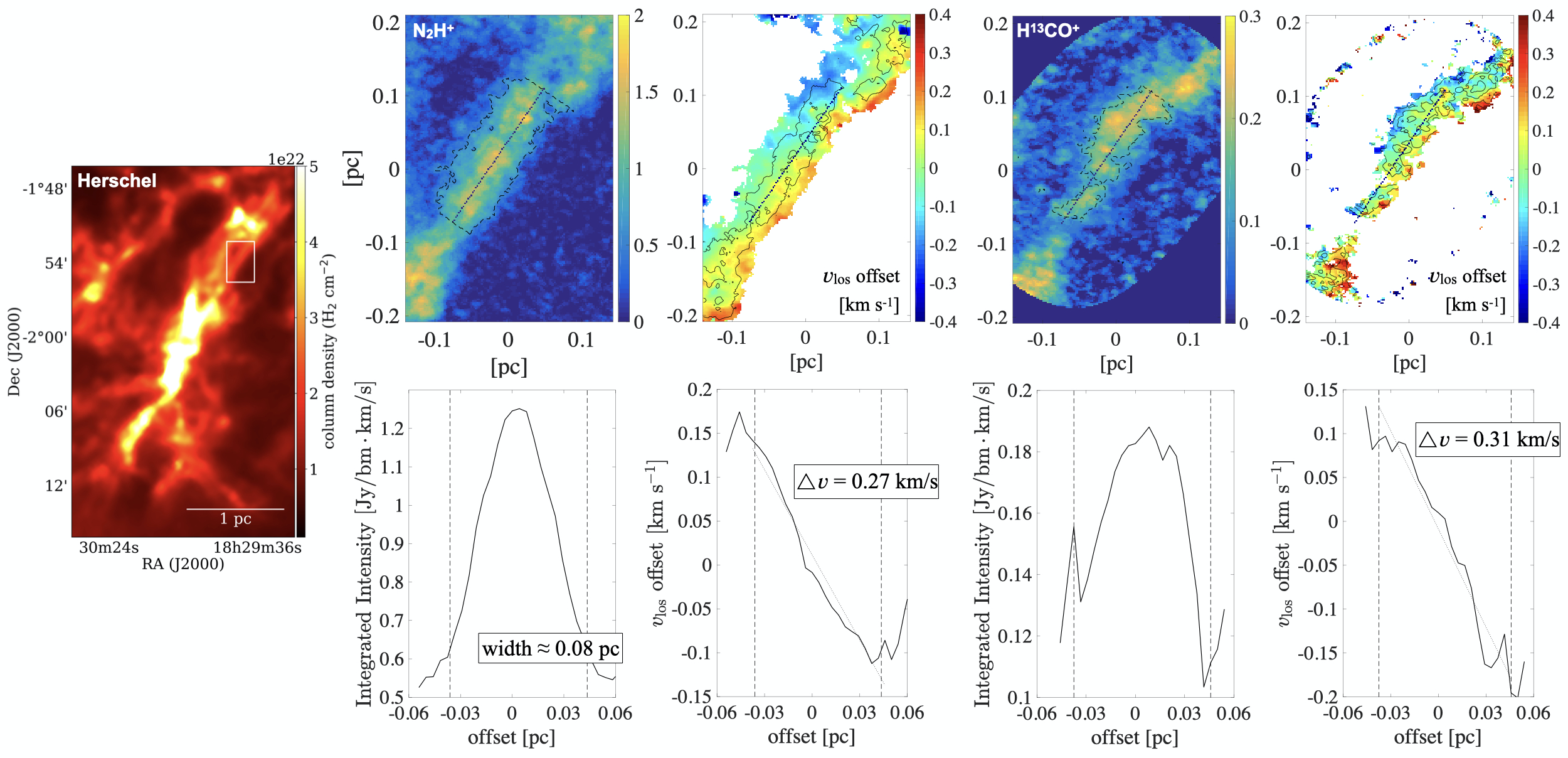}
%\vspace{-.1in}
\caption{{\it Left}: The {\it Herschel} column density map of Serpens South region, with the location of the Serpens South NW filament marked by white box. {\it Middle set of four panels}: The \NtwoH{} integrated intensity in \Jybm\kms{} ({\it top left}) and line-of-sight velocity offset ({\it top right}, with integrated intensity contours) maps of the SSNW filament from CLASSy, with the corresponding spatial profiles ({\it bottom sub-panels}) measured with respect to the spine of the filament ({\it dashed straight line in the upper panels}). {\it Right set of four panels}: Similar to the middle four panels, but for \HthrCO{} data from CLASSy-II. Both data sets are clipped at $2\sigma$ \citep[see][]{2018ApJ...853..169D}. The width of the filaments (in both \NtwoH{} and \HthrCO) are defined by the FWHM of the \NtwoH{} integrated intensity because of the slightly higher noise in \HthrCO{} data. The velocity differences measured from both tracers at the same width are also labeled on the velocity profile plots. }
\label{obsFilament}
\end{center}
\end{figure*}

\begin{figure}
\begin{center}
\includegraphics[width=\columnwidth]{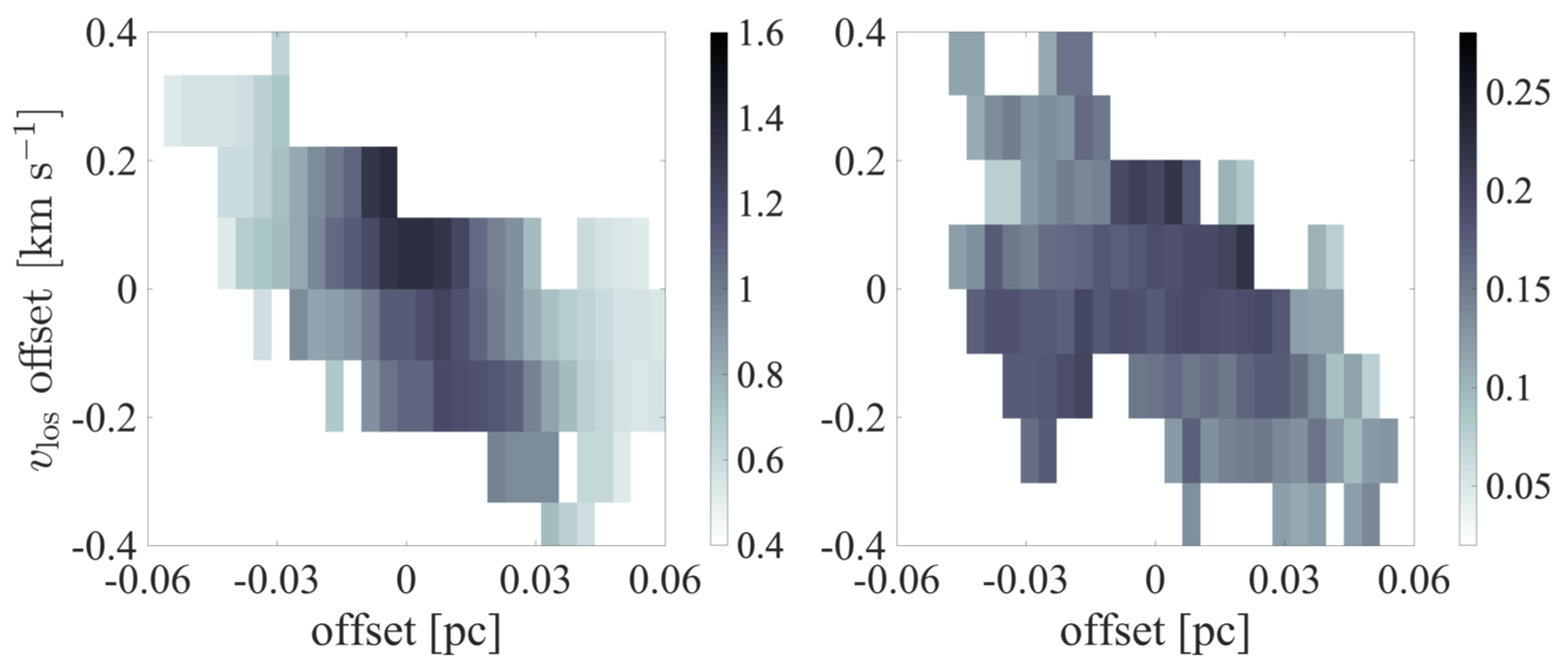}
%\vspace{-.1in}
\caption{The PV diagrams of the SSNW filament, from CLASSy \NtwoH{} data ({\it left}) and CLASSy-II \HthrCO{} data ({\it right}).}
\label{obsFilamentPV}
\end{center}
\end{figure}

In this work, we consider the \NtwoH{} data from CLASSy and the \HthrCO{} data from CLASSy-II of the Serpens South northwest (SSNW) filament as a real-life example of our filament formation model. 
The location of SSNW filament (northwest of the Serpens South hub) is marked by the white box in the left-most panel of Figure~\ref{obsFilament}, which shows the {\it Herschel} column density map of the Serpens South region.
%Other CLASSy filaments with similar radial veloctiy gradients are closer to hubs of star formation, making their structure and kinematics more complex---those filaments will be analyzed in upcoming work.
Note that there are other filaments in the Serpens South region showing similar velocity gradients \citep{2014ApJ...790L..19F,2018ApJ...853..169D}, but those filaments are closer to the bright, massive Serpens South hub of star formation, making their structure and kinematics more complex.
%and thus environmental effect on gas velocity structures cannot be simply ruled out. 
The SSNW filament is more isolated and away from the main star formation activity in the region, and therefore is ideal for our investigation here.
%\footnote{For a full CLASSy map of Serpens South, see Fig. 1 in \cite{2014ApJ...790L..19F}.}. 

%Figure~\ref{observedFilamentA} shows \NtwoH{} emission from the SSNW filament across six velocity channels. Using the ``x'' as a reference in the figure, it is clear that the SSNW filament emission moves from the southwest to the northeast between 7.6--7.2~\kms{}.  We model the \NtwoH{} emission profiles as Gaussians to derive the best-fit centroid velocity and line-of-sight velocity dispersion of the gas along the filament; we use the same methods detailed in \cite{2014ApJ...794..165S}. 
The upper-half of the middle and right four panels of Figure~\ref{obsFilament} summarizes the results from CLASSy \NtwoH{} and CLASSy-II \HthrCO{} observations towards the SSNW filament, showing the integrated intensity (left) and best-fit centroid velocity offset of the gas (right), which have been reported in \cite{2018ApJ...853..169D} (see their Fig.~2). 
The velocity offsets are calculated with respect to the median velocity of the gas within the filament boundaries (see below), $\langle v_\mathrm{fila} \rangle= 7.53$ and $7.54$~\kms{} for \NtwoH{} and \HthrCO, respectively.
%It is clear that the SSNW filament emission moves from the southwest to the northeast between 7.6--7.2~\kms{}
%Figure~\ref{obsFilament} shows the best-fit centroid velocity of the gas in the filament, making the radial veloctiy gradient even clearer. 
It is clear that the line-of-sight velocity of gas within the SSNW filament gradually shifts from $\sim +0.4$~\kms{} on the southwest edge to $\sim -0.1$~\kms{} on the northeast side in both \NtwoH{} and \HthrCO{} emissions.
This velocity gradient is better illustrated in Figure~\ref{obsFilamentPV}, which shows the PV diagrams of the SSNW filament from both \NtwoH{} (left) and \HthrCO{} (right) emissions. Here, the offset is calculated with respect to the spine of the filament (dashed straight line in the intensity maps; see below for definition). Both \NtwoH{} and \HthrCO{} emission distributions show steep slopes in the PV space; as discussed on the simulations in Section~\ref{sec::evo}, this is a clear evidence of velocity gradient across the filament.

%To quantify the magnitude, direction, and extent of the velocity gradient, we again need to define the boundary of the SSNW filament. However, 
Since neither of the \NtwoH{} and \HthrCO{} data successfully recovered a continuous spine of the SSNW filament (peak intensity at the center of the filament along its major axis), our method discussed in Section~\ref{sec::filaId} is not applicable here.
We therefore hand-picked the spine of the SSNW filament by eye, which is marked as a straight dashed line in the upper panels of Figure~\ref{obsFilament}. 
All pixels with $> 2\sigma$ signals in the centroid velocity map (regions within the dashed contour in the integrated intensity map) are included in calculating the radial profiles of integrated intensity and centroid velocity offset, which are plotted in the bottom-half of the four-panel set in Figure~\ref{obsFilament} for both \NtwoH{} (middle set) and \HthrCO{} (right set).
%The integrated intensity profile is then used to define the filament boundaries as when the integrated intensity drops to half of its maximum.
The filament boundaries are defined using the \NtwoH{} emission at where the intensity drops below half of the peak value.\footnote{This definition is different from that adopted in our simulated filament, which takes the ``background" into consideration (see Section~\ref{sec::filaId}). Since both \NtwoH{} and \HthrCO{} are considered as dense gas tracers, they should not be sensitive to lower density background gas.}
Note that because of the higher noise level of the \HthrCO{} data, the filament boundaries in \HthrCO{} uses those defined in \NtwoH{} data.
The two boundaries are marked by vertical dashed lines in the intensity and velocity profiles in Figure~\ref{obsFilament}; the width for the SSNW filament is therefore $D_\mathrm{SSNW} \approx 0.08$~pc.
%The width of the filament, or the FWHM of the \NtwoH{} integrated intensity, is therefore $\approx 0.08$~pc.

%We calculate the mean centroid velocity along the axial direction of the filament moving from the northeast edge of the filament to the southwest edge, and plot the results on the right of Figure~\ref{obsFilament}.  
To quantify the magnitude of the velocity gradient across the SSNW filament, we linearly fit the velocity profiles within the filament boundaries, and measure the velocity difference between the two ends of the fit, which is overplotted on the velocity profiles in Figure~\ref{obsFilament} (dotted lines).
The velocity difference across a width of $0.08$~pc in the SSNW filament is $0.27$ and $0.31$~\kms{} for \NtwoH{} and \HthrCO, respectively.
%The average velocity difference across the filament is 0.3~\kms. 
%As reported in \cite{2014ApJ...790L..19F}, there is no detected axial velocity gradient in this filament.

%To calculate a velocity gradient, we take slices across the width of
%the filament, fit a gaussian to the intensity profile, take the
%velocities at the FWHM, and subtract them.
%These two methods result in very similar answers of:

%(Original writing) The radial gradient in the SSNW filament is
%detected along a length of $\sim$0.3~pc and across a width of
%0.08~pc. 

%To calculate $C_v$ for the SSNW filament, we need to estimate its $M/L$ (see Equation~(\ref{coeff})).
%to go along with the velocity difference of 0.3~km/s across its width. s
%We assume that the filament is a uniform density cylinder, so that measurements of the cylinder radius and physical density directly give the cylinder $M/L$. To estimate the physical density, we use RADEX to model the ratio of \NtwoH{} J=3-2/J=1-0 line temperatures. The J=3-2 measurements were done using the Submillimeter Telescope (SMT) at the Arizona Radio Observatoy; we observed 12 Nyquist-spaced pointings along the spine of the filament, and smoothed the CLASSy data to the 30\arcsec{} beam of the SMT data. We compared the J=3-2 and J=1-0 spectra at each of the 12 pointings, and found a median H$_{2}$ number density of 1.8$\times$10$^{5}$~cm$^{-3}$. For a uniform cylinder, with $r$=0.04~pc, $n$=1.8$\times$10$^{5}$~cm$^{-3}$, and an assumed mean molecular weight of 2.37, the filament's $M/L$ is $\sim$61~\Msun~pc$^{-1}$. Therefore, this filament has $C_v$ $\sim$ 0.3.
From the {\it Herschel} column density map of the Serpens South cloud, we estimated an average column density along the SSNW filament to be $N_\mathrm{H} \sim 2.1\times 10^{22}$~cm$^{-2}$. Considering the filament width $D_\mathrm{SSNW} \approx 0.08$~pc, this yields the mass per unit length $M/L = N_\mathrm{H}m_\mathrm{H}\cdot D_\mathrm{SSNW} \approx 14$~\Msun~pc$^{-1}$. Therefore, the $C_v$ coefficient (see Equation~(\ref{coeff})) for the SSNW filament is about $0.36$. 
The fact that $C_v < 1$ in the SSNW filament indicates that the observed velocity gradient perpendicular to the filament could be induced by its self-gravity. 
In addition, this $C_v$ values agree with those derived from our simulated filaments for times when the filament is becoming prominent (see Figure~\ref{filaEvo}). We therefore propose that the SSNW filament is formed within a locally-flat region by gravitationally accreting material  anisotropically.  

\subsection{Subfilaments in PV Diagram: Multiple Structures or Density Selection Effect?}
\label{sec::density}

\begin{figure*}
\begin{center}
\includegraphics[width=\textwidth]{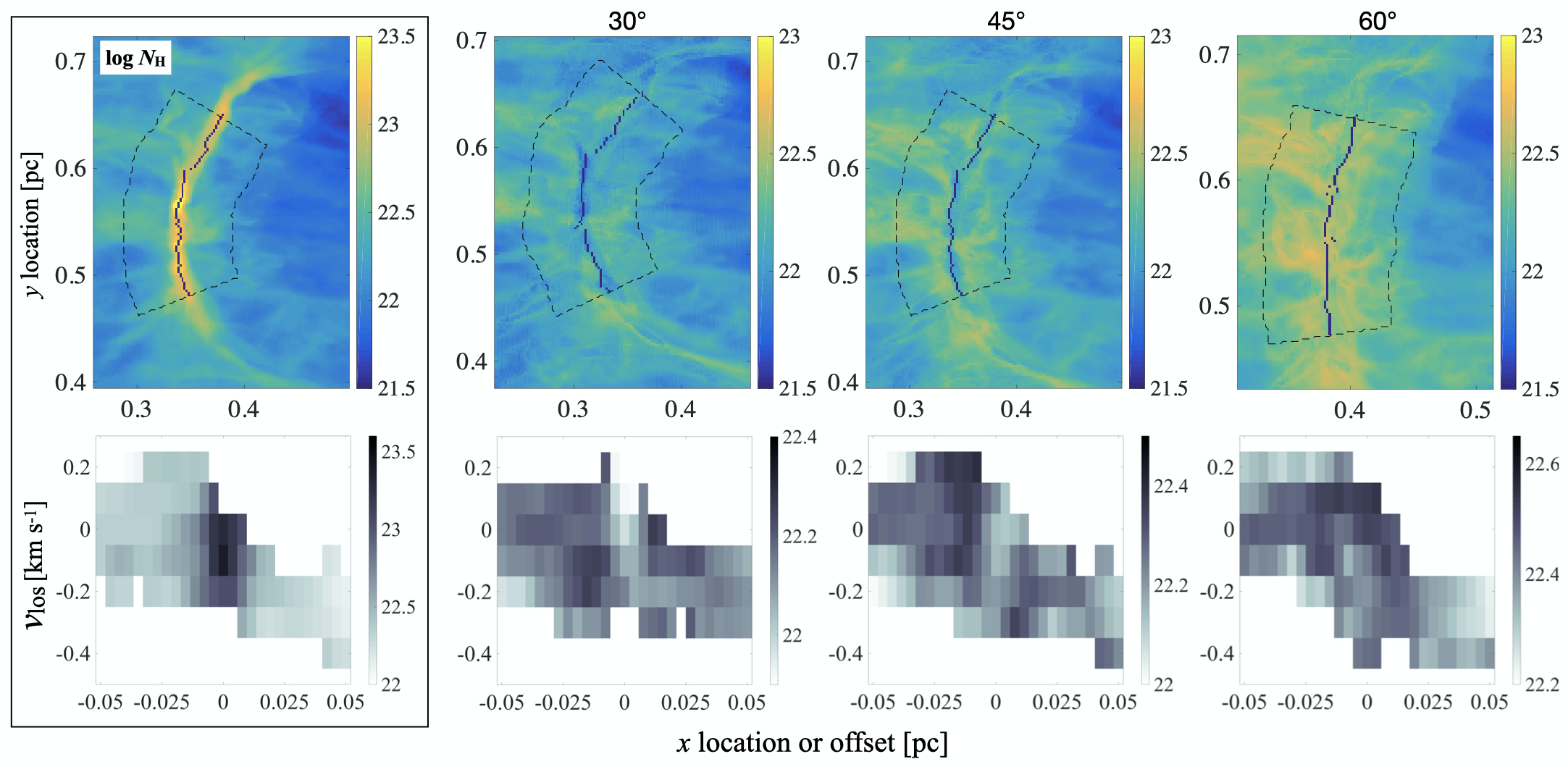}
\vspace{-.15in}
\caption{Illustration of the density selection effect at various viewing angle. The first column represents the reference case, which is from the last column of Figure~\ref{filaEvo}; i.e.~considering all cells with volume density with $10^{3.5} - 10^8$~cm$^{-3}$ with viewing angle $45^\circ$. Columns $2-4$ show the same filament viewed at different angles, and including only cells within the narrower density range $10^{3.5}-10^5$~cm$^{-3}$. {\it Top row:} $N_\mathrm{H}$ column density map (in $\log$ scale). {\it Bottom row:} PV diagram along the filament spine ({\it black lines} in the top row) for gas within the $0.05$~pc radius ({\it dashed regions} in the top row). The subfilaments can be clearly seen in the PV diagram for some viewing angles, which is simply caused by the missing high-density gas near the filament spine (small offset from the filament).}
\label{lowd}
\end{center}
\end{figure*}

%Figure~\ref{lowd} provides an explanation to the ``duplet" observed in \cite{2018ApJ...853..169D}. {\bf <<tbc>>}

\cite{2018ApJ...853..169D} found in their data an interesting feature that some filaments seem to break into two subfilaments in PV space (see e.g. their Fig.~9 and 12), with velocity difference $\sim 0.5-1.0$~\kms. Multiple velocity components have also been observed in the L1495/B213 filaments in Taurus \citep{2013A&A...554A..55H}, and have been interpreted as subfilaments which are either created from filament fragmentations \citep{2015A&A...574A.104T} or are going to collide to form more massive filaments \citep{2016MNRAS.455.3640S}. Here, as an extension of our filament formation model, we provide an alternative explanation to the subfilaments reported in \cite{2018ApJ...853..169D}: density selection effect.

Considering the nature of molecule excitation, each molecular line is prominent over a certain range of gas density. This means that for a given molecular line, it may not trace the entire filament all the way from the outer part to the central spine, which could have density (or column density) enhancement as large as an order of magnitude (see e.g. Figure~\ref{filaEvo}). 
A similar argument has been discussed in \cite{2018MNRAS.479.1722C}, who used synthetic C$^{18}$O observations of filaments formed in turbulent simulations to show that C$^{18}$O emission could be dominated by the outer envelopes of the central, overdense regions of the filaments and thus show multiple velocity components in the spectra.

Though detailed chemical modeling and radiative transfer calculation is beyond the scope of this paper, a simple experiment using our simulated filament (Figure~\ref{filaEvo}) demonstrates this {\it density selection effect}, which is illustrated in Figure~\ref{lowd}. 
The first column of Figure~\ref{lowd} shows the same column density map and PV diagram from the last column of Figure~\ref{filaEvo}, which represents the ``ideal" situation when every voxel in the filament is contributing equally to the integrated column density. These voxels have volume density range $n_\mathrm{H} \sim 10^{3.5}-10^8$~cm$^{-3}$.
With the same viewing angle ($45^\circ$ from the normal of the locally-flat shock-compressed layer where the filament formed), the third column of Figure~\ref{lowd} shows the integrated column density map and PV diagram of the filament with a density cutoff: only voxels with gas density within the range $n_\mathrm{H} = 10^{3.5}-10^5$~cm$^{-3}$ are included in calculating column density. 
Two separated components can be clearly seen in the resulting PV diagram,\footnote{Note that for cases with density cutoffs, the spine of the filament and the region considered in the PV diagram are derived from the original simulation data without density cutoff, because with density cutoffs the central part of the filament no longer corresponds to local maxima in column density, and therefore it is impossible to define the spine.} with velocity difference $\sim 0.3$~\kms. This nicely reproduces the subfilaments in PV space observed by \cite{2018ApJ...853..169D}.

However, one may ask why subfilaments in PV space are not always seen in observations, like the SSNW filament discussed in the previous section (see e.g. Figure~\ref{obsFilamentPV}). 
In addition to chemical dependence of molecular line emission on gas properties,
we argue that the viewing angle could also be critical in finding these subfilaments. In Figure~\ref{lowd} we present the column density maps and PV diagrams of the same filament with the same density cutoff, but from two additional inclination angles of the filament-forming layer with respect to the plane of sky, $30^\circ$ (second column) and $60^\circ$ (fourth column), to compare with the fiducial case $45^\circ$ (first and third columns). Obviously, the separation between the two components in the PV space becomes less prominent when the inclination angle is large (i.e.~when the layer is close to edge-on), and one can hardly distinguish the two ``subfilaments" in PV space for the case of inclination angle $=60^\circ$.
This can be easily understood as gas contents from the two sides of the filament overlap along the line of sight.

We would also like to point out that another feature of these density selection effect-induced subfilaments is that the spine of the filament is sometimes missing or less obvious in the integrated emission map, because there are not enough voxels along the line of sight at the central part of the filament that are within the required density range to contribute to the integration. In observations, this represents the case when the gas density around the filament spine is too high to emit the specific molecular line in consideration. This feature can be actually seen in some filaments reported in \cite{2018ApJ...853..169D} that have multiple components in PV space, like the Serpens South E filament (their Fig.~3) and the NGC\,1333 SE filament (their Fig.~6). 
%Looking at the \NtwoH{} integrated intensity maps of these two filaments, the brightest emissions appear to be offset from the central region at least for part of the filaments (e.g.~$\mathrm{RA} = 18^\mathrm{h}30^\mathrm{m}12^\mathrm{s} - 16^\mathrm{s}$ for Serpens South E filament and $\mathrm{DEC} = 31^\circ11$\arcmin$30$\arcsec$ - 12$\arcmin$30$\arcsec for NGC~1333 SE filament), in good agreement with our density selection model. 

Note that we do not claim that density selection effect is the reason for all observed subfilaments (or fibers) in velocity space, even though it does appear that these subfilaments are more common in moderate-density tracers (e.g.,~C$^{18}$O) than high-density tracers (e.g.,~\NtwoH). Future high-resolution continuum observations will be the key to revealing the unbiased gas structure of these filaments with multiple velocity components.

\section{Summary}
\label{discussion}

In this paper, we combine kinematics of observed and simulated dense gas structures to provide evidence for a model in which gas filaments form via in-plane flows parallel to the magnetic field in shock-compressed layers. CLASSy data towards Perseus and Serpens Molecular Clouds demonstrated prominent velocity gradients transverse to the filament major axis, a feature of the preferred-direction accretion model of filament formation, as shown in numerical simulations of \hyperlink{CO14}{CO14} and \hyperlink{CO15}{CO15}. The quantitative comparison between kinetic and gravitational energy also suggests that the observed velocity gradients are induced by the filament self-gravity. 

Our main conclusions are as follows:

\begin{enumerate}
\item We quantitatively examine a scenario to form dense, star-forming filaments within locally-flat gas layers compressed by supersonic turbulence within a molecular cloud. 
In the case that the shock is strong enough to compress gas to be  almost self-gravitating, local velocity perturbation within the shocked layer can lead to the filamentary structure by creating seeds that grow via gravitational collection of material. 
Within the layer, the gravity-induced accretion flows onto the forming filament are preferably parallel to the layer, which results in the kinetic feature of linear velocity gradient across the filament when viewed from any direction except face-on (Figure~\ref{formation}). This feature is commonly seen in numerical simulations (\hyperlink{CO14}{CO14}, \hyperlink{CO15}{CO15}) considering convergent flows that compress gas and form post-shock layers wherein filaments and prestellar cores develop  (Figure~\ref{gradientSample}).

\item To demonstrate that the velocity gradient perpendicular to the major axis of  a filament seen in our simulations is induced by the filament self-gravity, we follow the time evolution of a forming filament from one of our simulations (Figure~\ref{filaEvo}). The fact that the velocity gradient becomes more prominent with the filament unit mass (mass per unit length) supports our model of the origin of such velocity gradient. In addition, by quantitatively calculating the ratio between gas kinetic energy and filament gravitational energy $C_v$ (Equation~(\ref{coeff})), we show that the self-gravity of the filament is indeed strong enough to induce such velocity structure (Section~\ref{sec::evo}).

\item We test our model on a filament in the Serpens South northwest region (SSNW filament), which was firstly observed by CLASSy and later followed-up by CLASSy-II \citep{2018ApJ...853..169D}. Both the CLASSy \NtwoH{} and CLASSy-II \HthrCO{} data of the SSNW filament show velocity profiles in good agreement with our filament formation model very well, and with $C_v \sim 0.3$ the filament self-gravity is definitely responsible for the observed velocity gradient across the SSNW filament (Figure~\ref{obsFilament}).

\item As an extension of our filament formation model, we propose that the multiple components in PV space within an individual filament, as reported in \cite{2018ApJ...853..169D}, could be due to density selection effect. Instead of being real structures, these separated ``subfilaments" may simply reflect the fact that the central zone of the filament is too dense to emit certain molecular lines that are used to trace the outer part of the filament. We demonstrate this effect by adopting a simple density cutoff on synthetic observations toward one of our simulated filaments, and we note that this effect is dependent on the viewing angle with respect to the locally-flat, filament-forming layer (Figure~\ref{lowd}).

\end{enumerate}

\section*{Acknowledgements}

We thank the referee for a very helpful report. 
C.-Y.~C. is grateful for the support from NSF grant AST-1815784.
The work of E.C.O. was supported by grant 510940 from the Simons Foundation. 
CLASSy was supported by NSF AST-1139998 (University of Maryland) and NSF AST-1139950 (University of Illinois). Support for CARMA
construction was derived from the Gordon and Betty Moore Foundation,
the Kenneth T. and Eileen L. Norris Foundation, the James S. McDonnell
Foundation, the Associates of the California Institute of Technology,
the University of Chicago, the states of Illinois, California, and
Maryland, and the National Science Foundation. 
%Ongoing CARMA development and operations are supported by the National Science Foundation under a cooperative agreement, and by the CARMA partner universities.

%%%%%%%%%%%%%%%%% APPENDICES %%%%%%%%%%%%%%%%%%%%%

%\appendix


%If you want to present additional material which would interrupt the flow of the main paper, it can be placed in an Appendix which appears after the list of references.

%%%%%%%%%%%%%%%%%%%%%%%%%%%%%%%%%%%%%%%%%%%%%%%%%%

% Don't change these lines
\bsp	% typesetting comment
\label{lastpage}
\end{document}